# An Agnostic Domain Specific Language for Implementing Attacks in an Automotive Use Case

Christian Wolschke
Fraunhofer IESE
Kaiserslautern, Germany
christian.wolschke@iese.fhg.de

Stefan Marksteiner
AVL List GmbH
Graz, Austria
stefan.marksteiner@avl.com

Tobias Braun
Fraunhofer IESE
Kaiserslautern, Germany
tobias.braun@iese.fhg.de

Markus Wolf
AVL List GmbH
Graz, Austria
markus.wolf@avl.com

## Abstract

This paper presents a Domain Specific Language (DSL) for generically describing cyber attacks, agnostic to specific system-under-test (SUT). The creation of the presented DSL is motivated by an automotive use case. The concepts of the DSL are generic such that attacks on arbitrary systems can be addressed.

The ongoing trend to improve the user experience of vehicles with connected services implies an enhanced connectivity as well as remote accessible interface opens potential attack vectors. This might also impact safety and the proprietary nature of potential SUTs. Reusing tests of attack vectors to industrialize testing them on multiple SUTs mandates an abstraction mechanism to port an attack from one system to another. The DSL therefore generically describes attacks for the usage with a test case generator (and execution environment) also described in this paper. The latter use this description and a database with SUT-specific information to generate attack implementations for a multitude of different (automotive) SUTs.

## 1 Motivation

Testing systems against cyber attacks is typically performed for a specific system under test (SUT). The attacks contain knowledge about the potential structure of SUT. If other SUTs should be tested against these attacks, an adaption is required. To provide a means to generically describe known attacks in the needed manner, we developed the Agnostic domain-specific Language for the Implementation of Attacks: ALIA, which allows attack descriptions without encoding SUT-specific information. ALIA aims at the description of cyber attacks in a compact and easy understandable fashion. It has originally been developed according to the needs and requirements of the automotive domain for an attack language supporting a clear separation between the atomic *actions* (single steps of an attack) definition and SUT specific parts (e.g. bus configurations such as message IDs and formats). Despite its automotive roots, the resulting abstract concepts part of ALIA are indeed not automotive specific but can be used in different domains. Nevertheless, in this paper, we will describe its benefits and development aligned to an automotive use case. In the automotive domain, cybersecurity threats gain increasing importance due to enriched functionality by connected services which implies remotely accessible interfaces, as well as extended communication possibilities. External testers and certification authorities are supposed to test vehicles against potential attacks before permission for operation can be given. Furthermore the test results are fed back to system development to identify at which places additional security mechanism need to be added or existing needs to be extended. Based on the list of system functions and the knowledge of generic vehicle's architecture, analysis techniques like the security-aware hazard and risk analysis (SAHARA) method [10] can be applied to identify possible attack vectors and rank them by their critically (in terms of required efforts/knowledge as well as safety impact) already in early development phases. A threat library can help to systematically identify weaknesses and to develop implementation independent (abstract) attack descriptions. However, the translation of abstract attack descriptions into concrete executable attacks is a challenging task and requires the adaption to each vehicle's architecture. This adaption includes the consideration of the network structure and the available ECUs. The knowledge might either be given upfront by test engineers or the test system may inquire information online while executing the attack. In the first case, this knowledge is directly encoded in the test. There are two main drawbacks: the technical details complicate the test and its comprehensibility and hinder the reuse of the same test for different SUTs without manual adaption. Secondly, the maintenance of the same test for different platforms is time-consuming and error-prone when it comes to changes. The option to inquire information online is often rather complex, error-prone and might also create the need for adaption this inquiring mechanism to different SUTs. Furthermore, it extends the test time, might cause undesired side effects, and introduces a non-deterministic element that might thwart the comparability of tests. Therefore, experience leads to the conclusion that using raw attack scripts for concrete SUTs alone is not effective in terms of efforts and re-usability. Additionally, the executable attack implementations are hard to understand and their development requires in-depth knowledge of each targeted SUT. The goal of our approach is therefore to separate the SUT-specific information, as well as the used test environment from the pure attack description, while simultaneously reducing the complexity of developing attacks and the required in-depth knowledge of the SUT. This paper will provide an example, how an attack scenario is built up in Section 5, afterwards we will give an overview of the background of attack languages and explain why we have decided to develop our own solution (2). The following sections explain the concepts (3), the design (4) and implementation (6) of ALIA. The results are summarized in 7 and finally the conclusion is given (8).

## 2 Background

Several attack languages have been presented in the literature for different purposes [5]. It is pointed out that the testing may either focus on the correct implementation of a security mechanism or on the evaluation of vulnerabilities with risk-based testing approaches. Focusing on the vulnerability testing, it should be noted that they are often related to unintended side-effect behavior. The Meta Attack Language (MAL) uses the architecture encoded in a domain specific language to identify assets, their relationships and access techniques [8]. As language constructs, MAL chooses an object orientated approach, in which actions can be applied to certain assets. For each action, the next reachable actions can be given, so that attack paths can be built up dynamically. Furthermore, it allows to specify timeouts for actions. An application of the language to vehicular attacks is given by [9]. The MAL is well suited to identify potential attack paths. It allows for reasoning about assets and their potential attacks. The downside is that a complex system may lead to a huge variety of attacks and their impact to the risk of the system is unclear. Further, the need to model a specific attack on an asset and whether it may have an impact on the overall system might be hard to determine. Hence, we preferred explicit attack modeling for ALIA.

The application of natural language concepts to make ease-of-use in attack modelling is described by [11]. They propose pre- and postconditions per threat flow, which is a program containing if-clauses, loops and/or jump statements. Our ALIA approach took over these concepts, but extended the execution semantics so that even after failed statements the execution can proceed.

The ADeLe language [13] is a multi-purpose language for exploits, detection, correlation and response. It is also using pre- and postconditions to determine the feasibility of executing instructions and to evaluate the results. The action section, which actually describes the attack implementation, is not independent of the SUT.

Preconditions can also be formulated for each individual attack as shown in the Cyber-Physical Attack Description Language (CP-ADL) [15]. The notion of causes and effects allows to select follow-up actions.

The description of threats on a higher level of abstraction via use cases is proposed in [14]. The use case describes the step necessary to perform an attack. Pre- and postconditions describe when to apply the attack and which outcome is expected. This adopted the concept of pre- and postconditions of [13] and [14], but applied it to each action.

An alternative to text based attack descriptions are graphical notations. The CORAL approach uses sequence diagrams to identify potential attacks and to rate risk associated by considering the frequency of messages and the potential impact [3] [4]. The annotation of sequence diagrams with attack trees gives the possibility to model alternatives. The disadvantage of the language is its lack of formality. The language is intended to discuss potential attack scenarios and not to derive test cases automatically.

The specification of attack trees via a planning language is proposed by [2]. The idea is to specify actions that can be executed under certain preconditions. The effects of actions lead to new actions so that an attack tree gets created. They applied the Planning Domain Definition Language (PDDL) for implementation. The disadvantage of PDDL is that it assumes effects to be known in advance, which is not always the case of attacks. This would also make it hard to find an attack path to violate a safety goal, as the potential attack path have to be found by a search strategy. Our external approach will provide more scalability and flexibility than the current static parsing solution.

## 3 Concepts of DSL for attack descriptions

As outlined in Section 2, we have studied already existing concepts and evaluated the needs within our security testing department. Since most of the users do not have in-depth knowledge of all technical details of the SUT, we decided to store SUT specific aspects outside the attack descriptions. As pointed out in Section 1, this concept aims at improving re-usability, at reducing the complexity in development and at understanding these attack descriptions. Additionally, the test engineer should be able to focus on the attack itself and not on details of the SUT. Further requirements and design decisions derived from these considerations are:

- *Imperative based notation:* The imperative approach was pursued, as it is close to the existing notion of test scripts instead of developing a complex programming language able to describe attacks. As alternatives, also graphical notations were discussed, which allow for easily tracking cross-references between entities, but would not have a predefined fixed execution sequence of the single attack steps (called *actions*). Furthermore, attack trees were part of the discussion, as they could figure out at run-time which attacks could be executed; but they could not reveal easily which attack is planned for a system. While both concepts are interesting from an academic point view, they require changes in the established test process and training to be applied correctly, as well as additional efforts for modeling the inputs artefacts compared to the imperative approach.
- *Attack execution:* The attack should be consistent with actions, which each execute a certain attack tool and retrieve the necessary information. An action should group optional pre- and postconditions checked for each action, as well as the description of the action to be executed in this step. Within the specification of the command as well as the pre- and postconditions variables can be used. These variables are resolved during run-time using a vehicle database that contains vehicle specific data for the targeted SUT. If the preconditions are not met, the execution of the action will be skipped, as the preconditions are considered mandatory (sine qua non) and an execution is therefore not sensible. For an unfulfilled postcondition, a failed action will be logged, in order for a test engineer to know directly where and which steps failed. Pre- and postconditions can be linked (i.e. an unfulfilled postcondition could lead to an unfulfilled precondition for a subsequent step) and therefore allow conditional steps and additionally facilitate the interpretation of the test execution results (since steps with linked unfulfilled preconditions are not executed, see step 8 and 9 of Figure 1).
- *Failure semantics:* A usual way of checking the success of executions are assertions. Assertions can e.g. cause a program to terminate or to raise an exception if their condition is not met. In the DSL, the failing of a command would induce an exception within an action that should be caught by the test execution environment and the test should proceed with the next command. An example of the execution is given in Figure 1. It shows how the execution proceeds even after failures (as in Attack 1, Postcondition 1 or Postcondition 2). The failure in Precondition 3 leads to the skipping of Attack 3, but further actions are still be executed.
- *Predefined support for automotive bus systems:* Considering the use case, we decided, that the DSL should directly include support for testing automotive systems and their interactions/communication. Therefore, we directly integrated commands for commonly used communication buses used by the automotive industry (like CAN, FlexRay, MOST, etc.) as well as predefined message formats. Regarding the typical communication patterns, commands that consider the cyclic sending of messages as well as the specification of timeouts should be included.

| Time | Status | Debug message |
|---|---|---|
| 1 | OK | Precond 1 fulfilled |
| 2 | Failed | Attack 1 **failed** |
| 3 | Failed | Postcond 1 **failed** |
| 4 | OK | Precond 2 fulfilled |
| 5 | OK | Attack 2 executed |
| 6 | Failed | Postcond 2 **failed** |
| 7 | Failed | Precond 3 **failed** |
| 8 | skipped | Attack 3 not executed |
| 9 | skipped | Postcond 3 not checked |
| 10 | OK | Precond 4 fulfilled |
| 11 | OK | Attack 4 executed |
| 12 | OK | Postcond 4 fulfilled |

**Figure 1.** Execution semantic example

The goal described in Section 1 requires an abstraction layer in the test design, which we provide by introducing *test scenarios* as abstract counterparts of test cases. A single executable step (a *test script* that would e.g. use a specific exploit, scan for an interface or send a CAN message) inside this test scenario will have to be turned into a generic description, which we call a *test pattern*. These test patterns, that form a composite scenario, will have to be augmented with SUT information of the concrete SUT against which it is going to be used. This augmentation is part of the test case generation [12]. Figure 2 provides an overview of this process. This allows for porting attacks or test cases from one SUT to another, maintaining the attack's structure but unbinding it from the original SUT. Where specific information about the SUT or the test environment (e.g. connection of test bed to the SUT) configuration is required, placeholders can be used. Such a generic attack description is based on a sequence of actions, with optional pre- and postconditions. If an action fails in execution, the test should proceed with the further actions, so that other potential weaknesses of the SUT could still be detected (see Section 5). To obtain an executable attack implementation, we have implemented a generator and test execution environment, which takes the generic attack description (expressed in the developed DSL) and identifiers for the target SUT; as well as the test environment as an input. By querying a knowledge base containing specific information about the supported SUTs and test beds, the generic attack description is translated into an executable attack implementation for the selected SUT and test environment. Hence, a generic attack description specified in the DSL can be applied to different test beds and SUTs without modifications, while the generated executable attack implementation itself is executable without any further information. Tailoring the existing attack descriptions to new SUTs and test environment can be done by extending the database with new SUT-specific information or new information about the test environment. To proof and evaluate this approach, we have developed an integrated test case generator and execution environment along with the DSL using *Xtext* [6] and Python, respectively (see Section 3). This process of translating generic attack descriptions into concrete implementations could be referred to as turning abstract *test patterns* (similar to design patterns) into concrete *test scripts*. In the context of this paper a set of patterns is named *test scenario* and a set of attack implementations *test case* [12].

We have decided to use the *Xtext* [6][1], which is an Eclipse-based framework for developing DSLs.

The domain model and core concepts of our DSL have been specified as a meta-model before we started designing the concrete syntax of ALIA itself. Figure 3 shows an excerpt of this meta-model reduced to the core elements and their relations.

The core idea is that the generic *attack descriptions* written in ALIA are compiled into a representation in the JSON [7] data format. This *JSON representation* is interpreted by test execution processes (*AXE - Attack Execution Engine*). Thereby, system variables referring to SUT specific or test environment specific are resolved during execution by querying a provided *knowledge database*. For each executed test, the results of the actions are stored within a *test execution report* for evaluation.

The *attack description* written in ALIA consists of an arbitrary number of *attack steps* that can be used in multiple test executions. For the test execution, the attack step denotes which kind of *command* or which kind of control-flow action (i.e. *if-then-else* or *while* statement) is to be executed. The *command* is resolved by the *AXE* to a tool or script execution in the test environment. The *attack steps* may contain *preconditions* that asserted to determine if the step's execution is sensible and *postconditions* that are used evaluate the execution's success. *Labels* are used to link *preconditions* and *postconditions* to *actions*.

## 4 Design

Apart from the functional goals of an automotive attack DSL, our design targets towards allowing attacks to be written quickly but even more to enable an easy read- and reusability of the resulting scripts. To achieve this, the main attribute is the grouping of attack steps into preconditions, actions and postconditions (for the semantics see Section 3). Each generic attack description consists of a sequence of actions (i.e. atomic attack steps), which may link to pre- or postconditions via common labels. Variables can be either system variables which are resolved at runtime by the vehicle information available, or they can be auxiliary variables which store results of tool executions for the evaluation in further post- or pre-conditions in subsequent actions. The purpose is that, in conjunction with descriptive function naming, an (security) expert could quickly and easily comprehend the attack described by the script, i.e. understand the type and steps of the attack by a quick look-over. Pre- and postconditions consist of simple assertions based on variables and their values. Examples for such conditions are, the availability of a specific remote interface during execution of the attack on the SUT or the presence of an open shell. Actions primarily contain two types: *scan* and *exploit*. The former type is used for reconnaissance and targets towards finding vulnerable system components (interfaces, ports, APIs, etc.), while the latter uses the results of scans to execute the attack on the (potential) vulnerable systems. If the attack has been successful, the result is returned (e.g. an open shell on the target system). For flow control, ALIA also contains Boolean, arithmetic and comparison operators. The actions themselves are either of type *scan* that is used for reconnaissance purposes (e.g. to find a Bluetooth target) or type *exploit* that is used for manipulating systems (e.g. issuing exploit code, executing commands, etc.). The results of these types can be written on variables. Furthermore basic flow control (conditionals, loops) are part of the language. ALIA has been implemented in Xtext (example Listing 1) with a translation into a representation in the JSON data format (Listing 5). This translation step has been realized with Xtend [1]. The generated editor for ALIA has built-in support for generating embedded Eclipse editors providing syntax highlighting and checking. The implementation also generates warnings if system variables are unknown or any auxiliary variables are not defined before used (e.g. in the example, the misspelled variable `uesr` is detected and the error highlighted). This tool supports enables an easy and efficient way to develop attacks using ALIA.

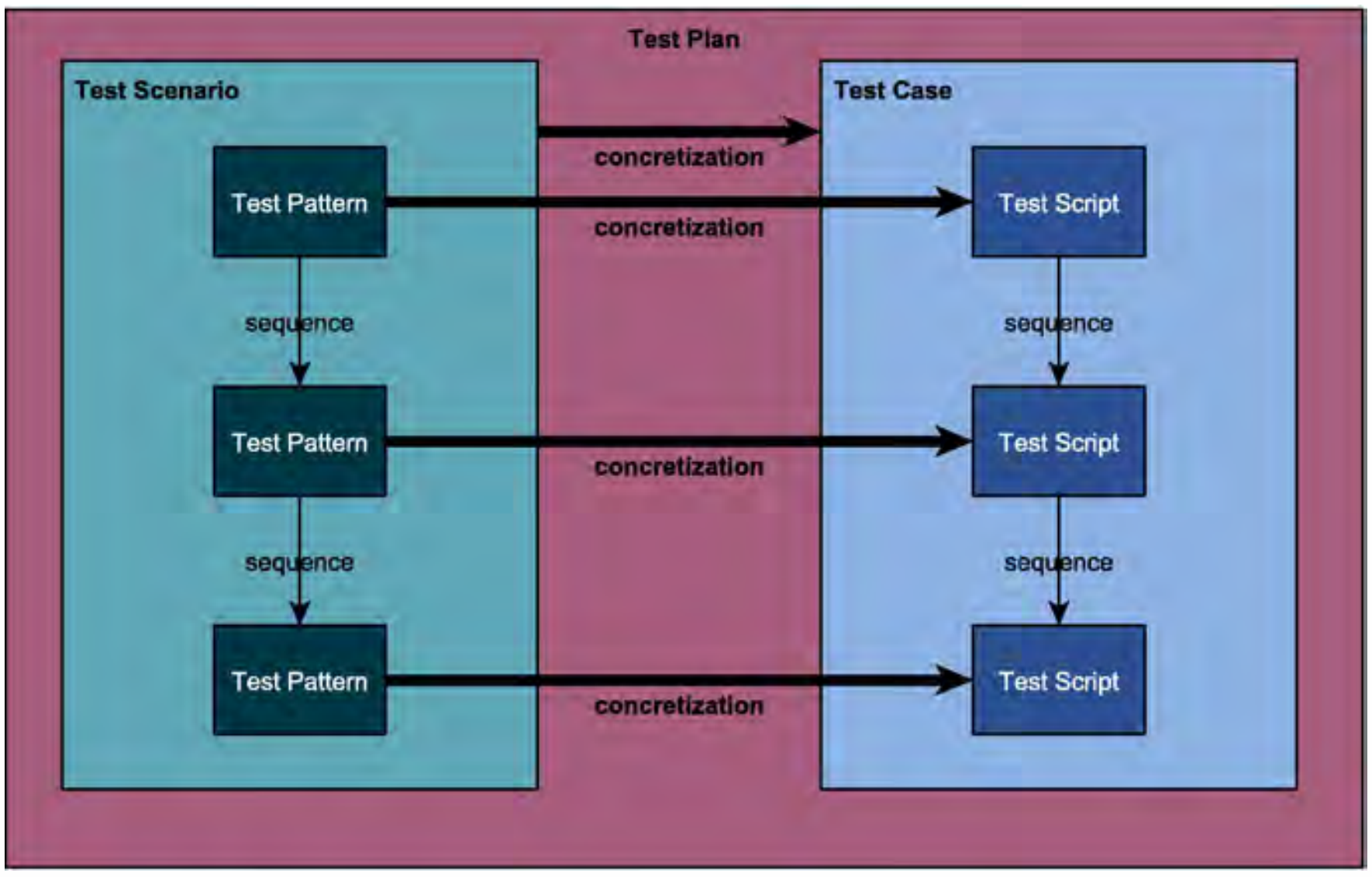

**Figure 2.** Abstraction and concretization concept from [12]

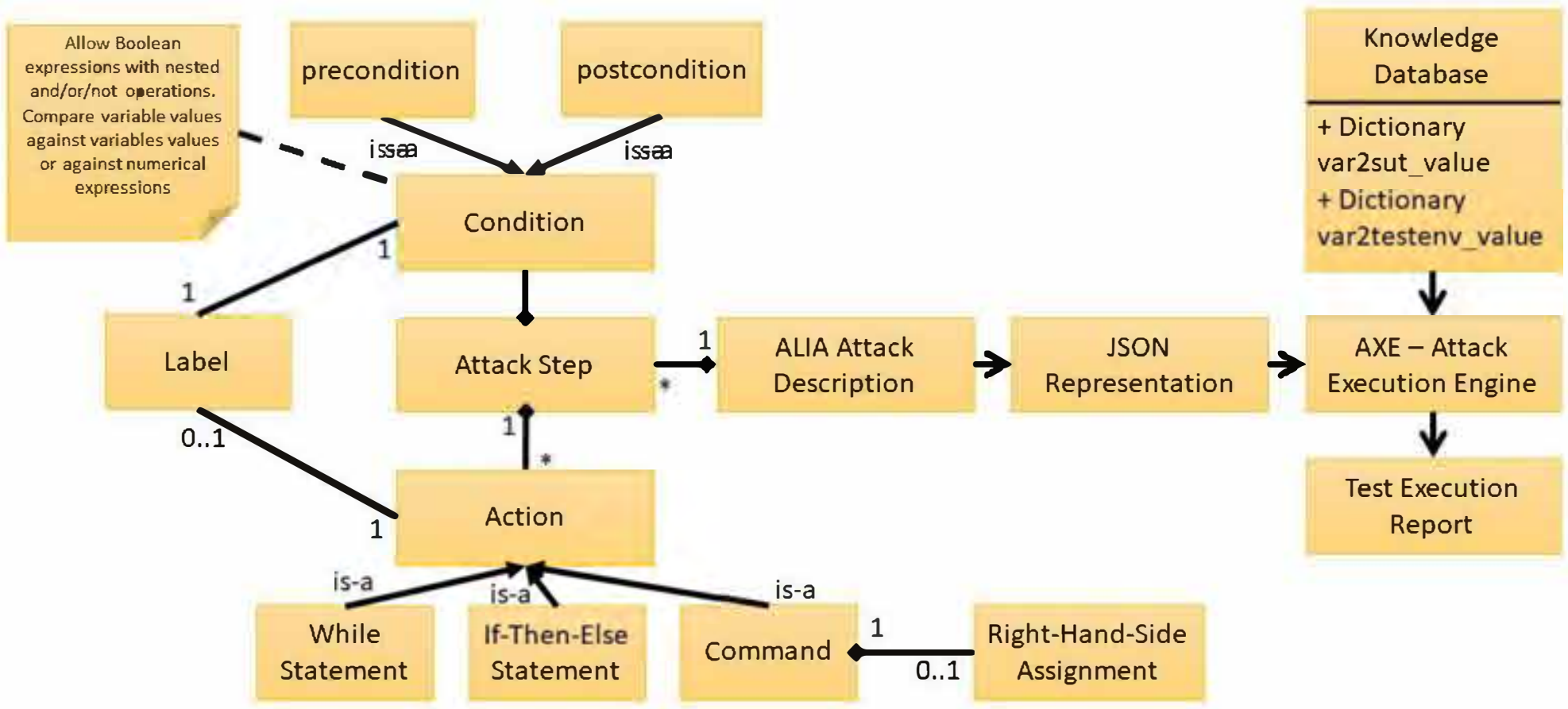

**Figure 3.** Meta-model of the ALIA

**Listing 1.** Automatic detection of unknown variables with ALIA

```
1 PreConditions:
2   get_su_rights: con
3 Actions:
4   get_con: con = exploit(type:OpenADB,
        target: ip_addr)
5   get_su_rights: exploit(type:
        ScriptExecution, command:'su')
6   exe_whoami: user = exploit(type:
        ScriptExecution, command:'whoami')
7   list: exploit(type:ScriptExecution,
        command:'ls')
8 PostConditions:
9   exe_whoami: uesr == "root"
```

Listing 1 shows an ALIA example attack description. The first action tries to open a connection to the SUT via the Android Debug Bridge (adb_connect; line 4); the required IP address for the connection is expressed via a variable and will be resolved at runtime. The warning in the editor in line 4 (expressed by the underlining) indicates, that the system variable ip_addr is not known to the test environment. Before executing line 5, the precondition associated with the label get_su_rights is checked. In the example it is required that the execution of line 4 was successful. Therefore, the precondition in line 2 asserts that the variable con contains an actual result and is not false before executing the action labeled get_su_rights. If the precondition is fulfilled, the system tries to acquire super user rights by executing the action in line 5. The result of the action is stored in the variable `user`, which is then analysed[1] using the postcondition of this action (line 9) against the word "root". Finally, the command "ls" is executed (line 7). As a summary, the composition of the language is as follows: Each action has a label as an identifier and a function (i.e. *scan* or *exploit*), which is of a certain type and takes certain input parameters (e.g. target, shell, etc.). The result may be stored to a variable. Each action can optionally have a precondition and/or a postcondition which both are attributed to the action by using the same label. Preconditions ordinarily assert the presence of a certain asset (e.g. an interface or a target) and cause an action to be skipped if not met. Postconditions contain expected results after the action (e.g. values a variable should have or expected measurements) and are used to evaluate the success of an action.

## 5 Example Use Case

As first use case ALIA has been applied to an automotive setting, in which an attacker wants to penetrate a vehicle. The flow of the specific attack (illustrated in Figure 4) was to attack a vehicle's infotainment head unit via a Bluetooth attack (1) and then open an Android Debug Bridge (ADB) (3) via WiFi to gain root access to the device. As potentially path an attacker can indirectly first attack a user's mobile phone (1,2), which is used as a trusted WiFi hotspot for updating and streaming. Once connected with elevated privileges, the head unit was used to send messages to the connected (via an USB tin or, alternatively, a Bluethooth OBD dongle) Controller Area Network (CAN) bus that contain fake speed and RPM values (4), which eventually manipulated the speed and RPM gauges. The (visible) result was that on a standing vehicle with active ignition (but still inactive motor), the respective gauges deflected. It is planned to keep such an attack agnostic, so that it can be applied also to other vehicles. For the given example, this would mean that the attack may need to try out several Bluetooth attacks to get the access to ADB connection. Even if this part of the attack would fail, we would like to have means which allow to continue the test. The argumentation is that the test may figure out whether multiple security controls prevent the attack to happen, or if only one security control actually prevents the attack. When the attack is proceeded with control over the Infotainment Head Unit the attacker would require to figure out, which busses can be accessed and also which messages may be sent to provoke a certain vehicle behavior. Based on this rationale, we derived the following requirements regarding the execution semantic: In order to adapt the test behavior during execution, it should be possible to check pre- and/or postconditions for each action. If the preconditions for executing a certain action are not met, the step would be skipped, and the execution would resume with the next action. This ensures, that if an action fails, the remaining attacks scripts can still be executed. Hence allowing for detecting otherwise not identified weaknesses which might be abused. The case that only one action failed in test execution, but the main goal could be reached, might indicate that only one single mechanism is effective. This would be worth to investigate further and take this into consideration for potential design changes. The formulated post-conditions allow for automatically evaluating the outcome of each action and, thereby, help in test automation.

## 6 Implementation and Test Case Generation

In order to generate an executable test case out of an attack description (test scenario), a JSON representation is compiled out of the attack (see example Listing 5). The compilation process is implemented in Xtend, which is a language that is completely inter-operable with Java. Due to such features as lambda expressions, dispatch methods, extension methods and type interference as well as multi-line template expressions, it is deemed an appropriate choice for programming code generators [1]. As each attack script consists of the three blocks preconditions, actions and postconditions, each line has to be identified and handled accordingly. Inside these blocks, every line consists of either a condition or a function call, which may be stored to a variable for further use. All pre- and postconditions are stored in a hash map structure, with the label as key and the corresponding command as value (see also Listing 3). In this process, for each command of the action section, the corresponding pre- and post-conditions are looked up via the label and placed before or after the corresponding command. Hence, conceptually and action consists of optional pre- and postconditions and the attack representing the executed command.

Listing 2 shows the parsing process for function calls. Depending on which function is used, the corresponding JSON formatted text is added to the output and introduced placeholders are replaced with actual values if they are present in the ALIA script or with placeholders that are used during the execution, if they have to be determined at runtime.

The example in Listing 2, written in Xtend, shows also how the commands of the DSL are translated to the naming of tools and parameters. Please notice, that the variable `ip_addr` gets translated to the actual value (e.g. "192.168.1.1") via the vehicle database containing the SUT specific details, which is processed together with the JSON file at the test's runtime (for this value is not known a priori and may change depending on the SUT).

The generator shown in Listing 3 is used to parse a previously defined ALIA script into a JSON output script using the defined transformation in Xtend. It adds a skeleton structure to produce a valid JSON definition and then reads and compiles the input script line by line from the source.

[1]For demonstration purposes, we misspelled the variable here to demonstrate the variable checking support of the editor.

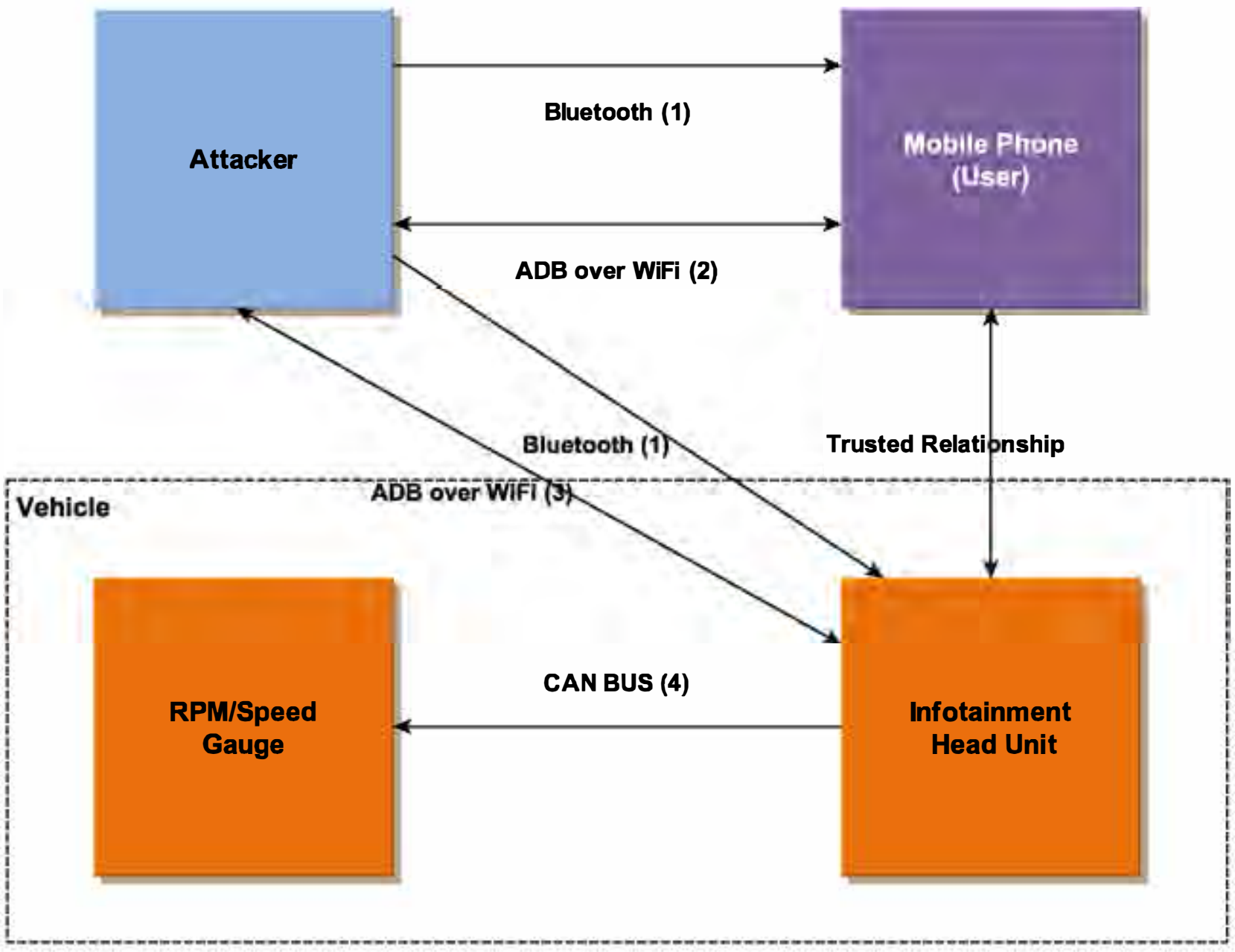

**Figure 4.** Example Attack Schematics

## 7 Experimental Results

In order to put the DSL ALIA to a practical test, the language was applied to the specific use case presented in section 5.

This attack should be automatically executed on a test system. Therefore, the attack was modelled in ALIA and subsequently turned into a system-interpretable representation using Xtend for test case generation. Listing 4 displays the resulting (to some extend anonymized) description in ALIA. The Xtend rules generate an executable attack script out of this description, using a repository of exploit scripts (e.g. in line 14), executables and a vehicle database. This vehicle database contains, for instance, SUT-specific concrete values for CAN messages, for which the ALIA description holds generic identifiers (e.g. MSG_SPD, might translate to 123#ABCD000000 for one SUT but 200#CAFE123456 for another). In our proof-of-concept setup, the generated attack script as JSON file is later on used as input for the AXE, which is a platform independent security testing application programmed in Python. Its purpose is to execute the commands from the attack script against the SUT and gather feedback from it. The generated JSON file from the DSL is taken as Input via a HTTP request and every command gets executed line by line. Furthermore it manages different shell connections and environments automatically, which allows for example to execute an exploit in a bash shell and then use the resulting reverse shell for further commands. Afterwards, the initial shell may be used again. Depending on the pre- and postconditions of the DSL script, some of the commands in the attack can be skipped and all output data from executed commands is collected into the response, which is sent back via a HTTP response. In the proof-of-concept, a malicious CAN message is periodically sent to CAN bus of the SUT (in this case a Mazda Model 3 from 2012) with the content "201#32C800006464C800". This message lets the instrument cluster of the vehicle assume that it is moving at maximum speed and rpm and therefore it starts to move its needles.

Listing 5 shows a snipped from a generated JSON attack script for this example. The execute block contains a sequence of commands. Each command has an environment and tool and a list of parameters. Besides actions, pre- and post- condition checks are also translated to the call of commands. The parameters may contain variable values, as `{ip_addr}`, which are to be replaced by the actual value by the AXE, taken from the vehicle database. These generated instructions are sufficient for the AXE to actually execute the attack as a fully sequenced test case.

## 8 Conclusion and future work

This paper presents the concept, design and implementation of an SUT-agnostic attack language ALIA, that allows for describing attacks on automotive systems in a generalized manner. ALIA separates the SUT specific information from

**Listing 2.** Xtend Parsing function of the output generator

```
private def compile_func(FuncCall e)
'''IF (e.f1000 !== null){"environment":"
    bash", "parameters":["connect", "{e.
    ip_addr.name}"], "tool": "adb"} ENDIF
IF (e.f1001 !== null){"environment":"bash
    ", "parameters":["connect", "{e.
    ip_addr.name}"], "tool": "e.intf"}
    ENDIF
IF (e.f1010 !== null)compile_f1010(e)
    ENDIF
IF (e.f1011 !== null)compile_f1011(e)
    ENDIF
IF (e.f1020 !== null)compile_f1020(e)
    ENDIF
IF (e.f1030 !== null)compile_f1030(e)
    ENDIF
IF (e.f1040 !== null)compile_f1040(e)
    ENDIF
IF (e.f1050 !== null)compile_f1050(e)
    ENDIF
'''

private def compile_f1010(FuncCall e){
  var retval = '{"environment":"adb"';
  retval += ', "parameters":[';
  retval += compile_stringformat(e.
      systemstr, compile_varlist(e.
      varlist));
  retval += ']';
  retval += ', "tool": "'+e.systemstr.
      split(" ").get(0)+'"}';
  return retval;
}
```

the test case. Hence, test cases can be applied to multiple different SUTs. ALIA is capable of abstracting from SUT specific tools and parameter values. The execution of ALIA scripts is made flexible by integration of pre- and post-conditions together with it's attack execution semantics. The first implementation of ALIA with Xtext [6] and the implementation of the attack execution are demonstrated. It is planned to use ALIA to implement attacks reflecting different scenarios for automotive testing system applicable to different SUT by providing the according information in the vehicle database. Future work is supposed to extend the provided functionality and usability of the ALIA. Usability features to be included into the code editor are automatic code completion and the integration of type system for variables. The provided functionality will be extended by adding support for additional tools (e.g. penetration and fuzzing tools). It is planned to use the ALIA as a common agnostic test specification for a multitude of different test vehicles. Plans for further enhancement of the ALIA include the integration of conditionals (IF-statements) and generalization of various different function categories, e.g. a central function for scans on different interfaces such as WiFi or BlueTooth. In this case, distinction will be made only by the provided parameters. Another improvement will be to implement data storage classes for used objects such as scans, exploits, interfaces or shell-connections and to load the different preconfigured flavours of these items from an external database. An entry in this database for an exploit may consist of an identifier, a description, required input parameters, output source code and a version number. This external approach will provide more scalability and flexibility than the current static parsing solution.

**Listing 3.** Source of the output file generation

```
class AttackDSLGenerator extends AbstractGenerator {
    @Inject extension IQualifiedNameProvider
    HashMapList<String, Precondition> precond_hashmaplist = new HashMapList<String,
        Precondition>();
    ArrayList<AttackStep> attack_list = new ArrayList<AttackStep>();
    HashMapList<String, Postcondition> postcond_hashmaplist = new HashMapList<String,
        Postcondition>();

    override void doGenerate(Resource resource, IFileSystemAccess2 fsa, IGeneratorContext
        context) {
    storeAst(resource, fsa)

    precond_hashmaplist = new HashMapList<String, Precondition>();
      attack_list = new ArrayList<AttackStep>();
      postcond_hashmaplist = new HashMapList<String, Postcondition>();

    for (e : resource.allContents.toIterable.filter(Line)){
      compile(e);
    }

      var retval = '{\n   "execute":[\n'
      if (attack_list.size()>=1){
        for (cmdctr: 0..<attack_list.size()){
          retval += compile_attack_list(cmdctr);
        }
        retval = retval.substring(0, retval.length() - 1);
        retval += "   ]\n}"
      } else {
        retval += "   ]\n}"
      }
      fsa.generateFile(
        resource.URI.lastSegment + ".json",
             //"test_script.py",
             retval)
    }
```

**Listing 4.** ALIA example of an automotive attack

```
PreConditions:
    bb_bt_scan: BT_IF
    bb_exploit: mytarget
    open_hotspot: bbshell
    install_python_lib: adbshell
Actions:
    bb_bt_scan: mytarget = scan(type:BlueBorne, interface:BT_IF)
    bb_exploit: bbshell = exploit(type:BlueBorne, target:mytarget)
    open_hotspot: mytarget.ip = exploit(type:OpenAndroidHotspot, target:mytarget, shell:bbshell)
    adb_con: adbshell = exploit(type:OpenADB, target:mytarget)
    install_python_env: exploit(type:InstallPythonEnv, target:mytarget)
    install_attack_script: attackScript = exploit(type:InstallAndroidCANDosScript, target:mytarget)
    install_python_lib: exploit(type:InstallPythonLib, target:mytarget, shell:adbshell)
    can_attack: exploit(type:ScriptExecution, target:mytarget, shell:adbshell, file:CarCanAttackScript)
PostConditions:
    open_hotspot: WIFI == "Android"
    can_attack: Oracle.CAN_MESSAGE(MSG.SPD)
```

**Listing 5.** Generated attack script

```
{
  "execute":[
    {"environment":"bash", "tool":"adb",  "parameters":["connect","{ip_addr}"]},
    {"environment":"adb", "tool":"was_sucessful",  "parameters":["label_con"]},
    {"environment":"adb", "tool":"su", "parameters":[]},
    {"environment":"adb", "tool":"whoami", "parameters":[]},
    {"environment":"adb", "tool":"python" "parameters":["condcmp.py","user","==","root"]},
    {"environment":"adb", "tool":"whoami", "parameters":["<>","\"root\""]},
    {"environment":"adb", "tool":"ls", "parameters":[]}
  ]
}
```